\documentclass{article}

\usepackage{graphicx}

\begin{document}

\title{Dark Matter and Dark Energy are mirages}
\date{October 4, 2017}
\author{G. Quznetsov \\ 
mailto: quznetso@gmail.com}
\maketitle

\begin{abstract}
In arXiv:hep-ph/9812339 the probabilities of point events in space 3 + 1 obey an equation of Dirac type .

Masses, moments, energies, spins, etc. are the parameters of the probability distribution of such events.

The terms and equations of quark-gluon theories turn out theoretically probabilistic terms and theorems.
Confinement and asymptotic freedom are explained by behaviour of such probabolities. And here we have the probabilistic foundations of the theory of gravitation.

In these article the Dark Matter and Dark Energy phenomenoms are explained by these results.

Knowledge of the elements of linear algebra and differential calculus is sufficient to understand the content of this article
\end{abstract}

\section{Introduction}

	In the study of the logical foundations of probability theory, I found that the terms and equations of the fundamental theoretical physics 
represent terms and theorems of the classical probability theory, more precisely, of that part of this theory, which considers the probability of 
dot events in the 3 + 1 space-time.

	In particular, all Standard Model's formulas (higgs ones except) turn out theorems of such probability theory. And the masses, moments, 
energies, spins, etc. turn out of parameters of probability distributions such events. The terms and the equations of the electroweak and of the 
quark-gluon theories turn out the theoretical-probabilistic terms and theorems. Here the relation of a neutrino to his lepton becomes clear, 
the W and Z bosons masses turn out dynamic ones, the cause of the asymmetry between particles and antiparticles is the impossibility of the birth 
of single antiparticles. In addition, phenomena such as confinement and asymptotic freedom receive their probabilistic explanation. And here we have 
the logical foundations of the gravity theory with phenomena dark energy and dark matter.

The proposed article contains initial concepts and the results of the development of these ideas.

\section{Dark Energy}

In 1998 observations of Type Ia supernovae suggested that the expansion of the universe is accelerating \cite{Ri}. 
In the past few years, these observations have been corroborated by several independent sources \cite{S}. This expansion is 
defined by the Hubble\index{Hubble, Edwin}\footnote{Edwin Powell Hubble (November 20, 1889 – September 28, 1953)[1] 
was an American astronomer} rule \cite{HB}:

\begin{equation}
V\left( r\right) =Hr\mbox{,}  \label{10}
\end{equation}

here $V\left(r\right)$ is the velocity of expansion on the distance $r$, $H$ is the Hubble's constant 
($H\approx 2.3\times 10^{-18}c^{-1}$ \cite{Ch}). 

Let a black hole be placed in a point $O$. Then a tremendous number of quarks\index{quark} oscillate in this point. These oscillations 
bend time-space and if $t$ has some fixed volume, $x>0$, and $\Lambda :=\lambda t$ then \cite[p.15,p.21]{QCD}

\begin{equation}
v\left( x\right) =\mathrm{c}\tanh \left( \frac \Lambda {x^2}\right)\mbox{.} 
\label{4}
\end{equation}

\begin{figure}
\centering
\includegraphics[width=0.75\textwidth]{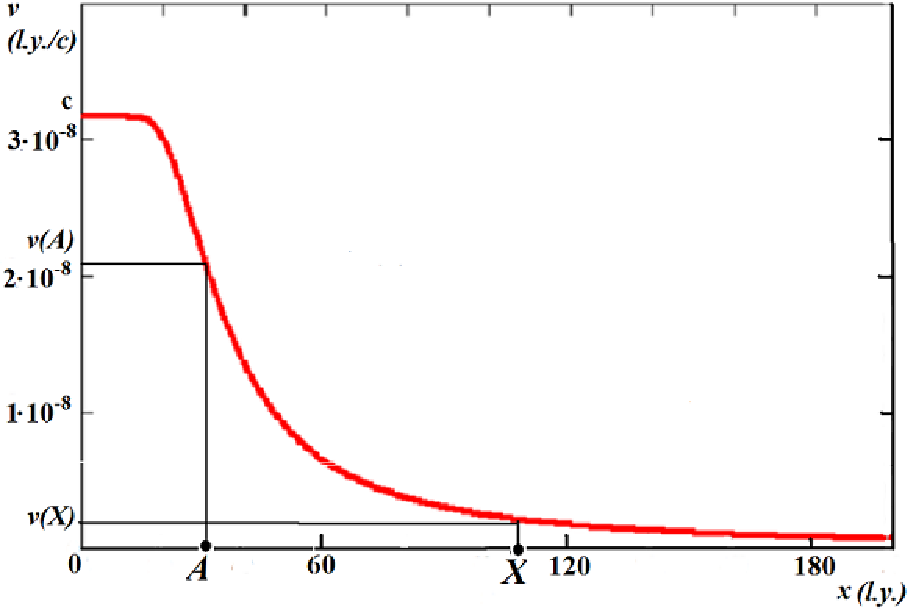}
\caption{Dependence of $v$ (light year/c) on $x$ (light year) with $\Lambda = 741.907$}
\label{Figure: 1}
\end{figure}

A dependency of $v(x)$ (light years/c) from $x$ (light years) with $\Lambda = 741.907$ is shown 
in Figure \ref{Figure: 1}. 

Let a placed in a point $A$ observer be stationary in the coordinate system $%
\left\{ t,x\right\} $. Hence, in the coordinate system $\left\{ t^{\prime
},x^{\prime }\right\} $ this observer is flying to the left to the point $O$ with
velocity $-v\left( x_A\right) $. And point $X$ is flying to the left to the point $O$ with velocity $-v\left(
x\right) $.

\begin{figure}
\centering
\includegraphics[width=0.75\textwidth]{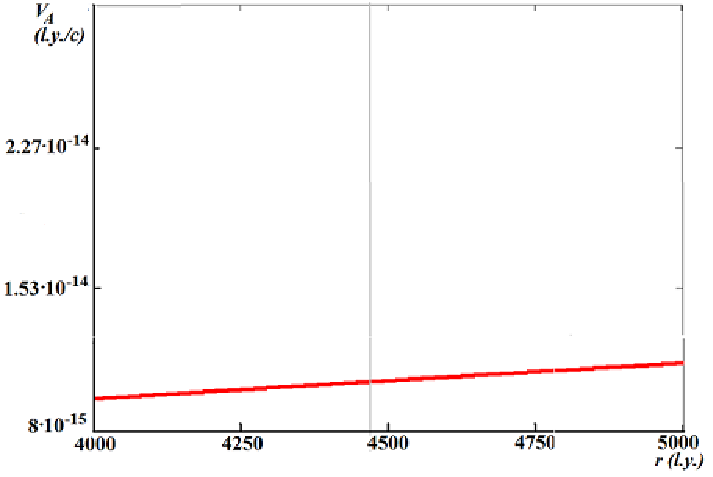}
\caption{Dependence of $V_A\left( r\right) $ on $r$ with $x_A=25\times 10^3$ l.y.}
\label{Figure: 2}
\end{figure}

\begin{figure}
\centering
\includegraphics[width=0.75\textwidth]{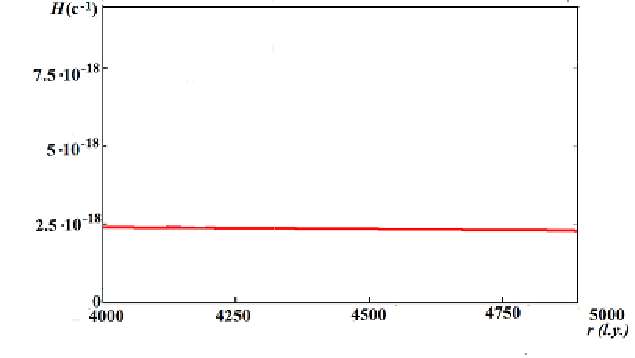}
\caption{Dependence of $H$ on $r$}
\label{Figure: 3}
\end{figure}

Consequently, the observer $A$ sees that the point $X$ flies away from him
to the right with velocity

\begin{equation}
V_A\left( x\right) =\mathrm{c}\tanh \left( \frac \Lambda {x_A^2}-\frac \Lambda
{x^2}\right)   \label{6}
\end{equation}

in accordance with the relativistic rule of addition of velocities.

Let $r:=x-x_A$ (i.e. $r$ is distance from $A$ to $X$), and

\begin{equation}
V_A\left( r\right) :=\mathrm{c}\tanh \left( \frac \Lambda {x_A^2}-\frac
\Lambda {\left( x_A+r\right) ^2}\right)\mbox{.}   \label{7}
\end{equation}

In that case Figure \ref{Figure: 2} demonstrates the dependence of $V_A\left( r\right) $
on $r$ with $x_A=25\times 10^3$ l.y.

Hence, $X$ runs from $A$ with almost constant acceleration:

\begin{equation}
\frac{V_A\left( r\right) }r=H\mbox{.}  \label{8}
\end{equation}

Figure \ref{Figure: 3} demonstrates the dependence of $H$ on $r$. (the Hubble constant.).

Therefore, the phenomenon of the accelerated expansion of Universe is explained by oscillations of chromatic\index{chromatic} states.

\section{Dark Matter}

"In 1933, the astronomer Fritz Zwicky%
\index{Zwicky, Fritz}\footnote{%
Fritz Zwicky (February 14, 1898 -- February 8, 1974) was a Swiss astronomer.}
was studying the motions of distant galaxies. Zwicky estimated the total
mass of a group of galaxies by measuring their brightness. When he used a
different method to compute the mass of the same cluster of galaxies, he
came up with a number that was 400 times his original estimate. This
discrepancy in the observed and computed masses is now known as "the missing
mass problem." Nobody did much with Zwicky's finding until the 1970's, when
scientists began to realize that only large amounts of hidden mass could
explain many of their observations. Scientists also realize that the
existence of some unseen mass would also support theories regarding the
structure of the universe. Today, scientists are searching for the
mysterious dark matter not only to explain the gravitational%
\index{gravitation} motions of galaxies, but also to validate current
theories about the origin and the fate of the universe" \cite{Miller}
(Figure 8 \cite{Hawley}, Figure 9 \cite{Begeman}).

\begin{figure}[htbp]
\centering
\includegraphics[width=.75\textwidth]{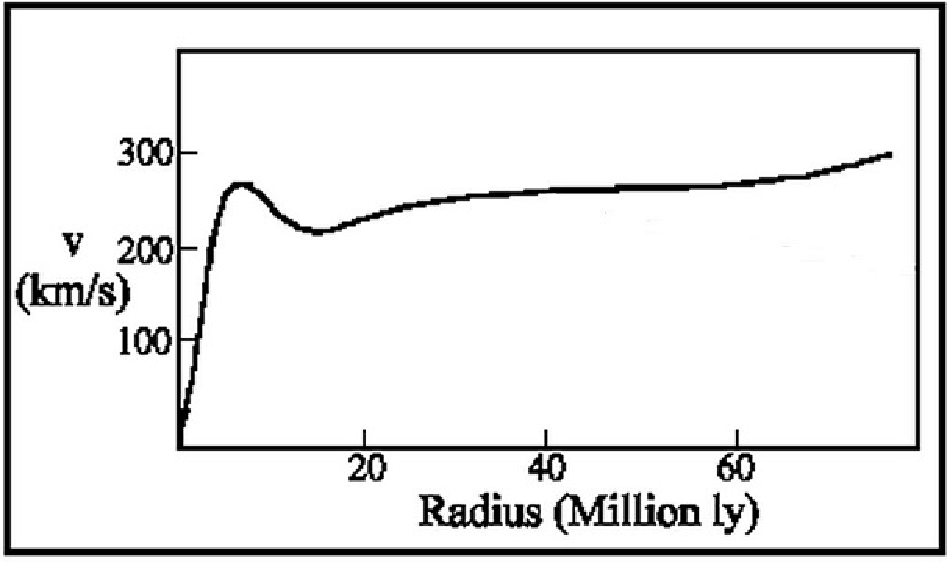}
\caption{A rotation curve for a typical spiral galaxy. The solid line shows actual measurements (Hawley and Holcomb., 
1998, p. 390) \cite{Hawley}}
\end{figure}

\begin{figure}[htbp]
\centering
\includegraphics[width=.75\textwidth]{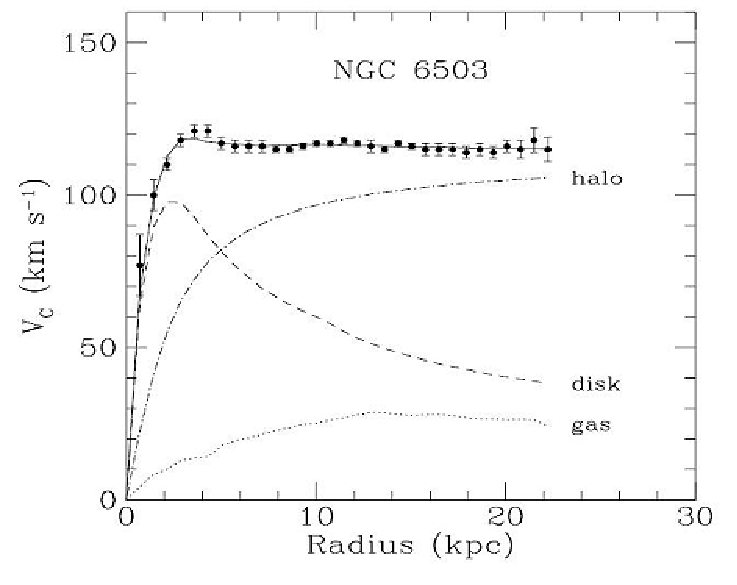}
\caption{Rotation curve of NGC 6503. The dotted, dashed and dash-dotted
lines are the contributions of gas, disk and dark matter, respectively.}
\end{figure}

\begin{figure}[htbp]
\centering
\includegraphics[width=.75\textwidth]{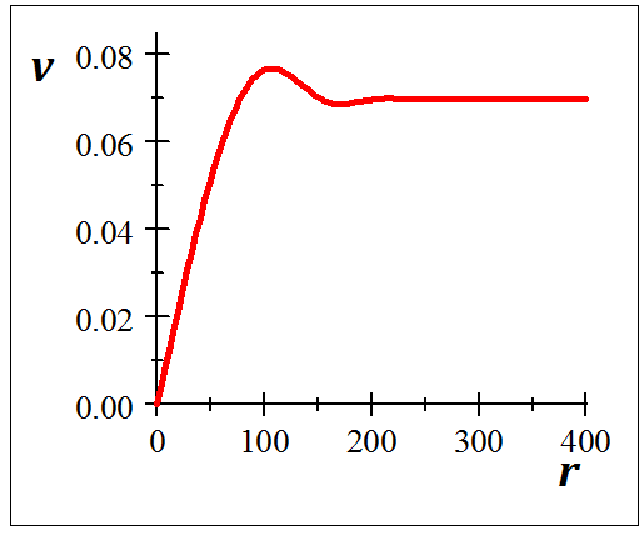}
\caption{For $t=10000$, $\theta =13\pi/14$:}
\end{figure}

\begin{figure}[htbp]
\centering
\includegraphics[width=.75\textwidth]{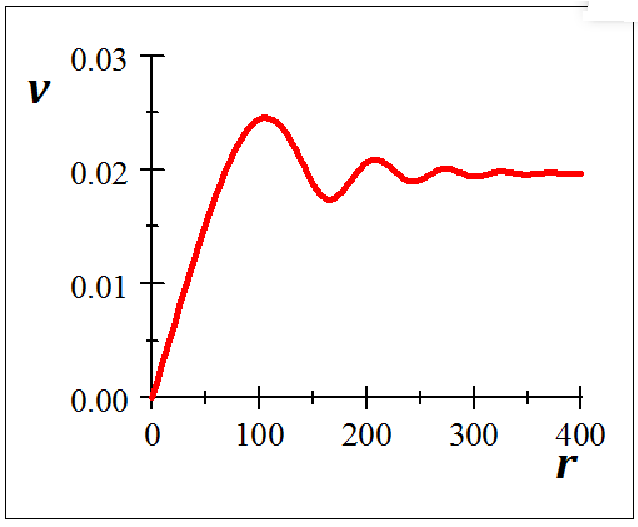}
\caption{For $t=10000$, $\theta =0.98\pi $:}
\end{figure}

Some oscillations of chromatic%
\index{chromatic} states bend space-time as follows \cite[p.~13]{QCD}  : 
\begin{eqnarray}
\frac{\partial }{\partial x^{\prime }} &=&\cos 2\alpha \cdot \frac{\partial 
}{\partial x}-\sin 2\alpha \cdot \frac{\partial }{\partial y}\mbox{,}
\label{u7} \\
\frac{\partial }{\partial y^{\prime }} &=&\cos 2\alpha \cdot \frac{\partial 
}{\partial y}+\sin 2\alpha \cdot \frac{\partial }{\partial x}\mbox{.}  \nonumber
\end{eqnarray}

Let

\begin{eqnarray*}
z &:&=x+\mathrm{i}y, z=re^{\mathrm{i}\theta }\mbox{.}\\
z^{\prime } &:&=x^{\prime }+\mathrm{i}y^{\prime }\mbox{.}
\end{eqnarray*}

Because linear velocity of the curved coordinate system $\left\langle
x^{\prime },y^{\prime }\right\rangle $ into the initial system $\left\langle
x,y\right\rangle $ is the following \footnote{$x^{\prime \bullet }:=\frac{\partial x^{\prime }}{\partial t}$, $y^{\prime
\bullet }:=\frac{\partial y^{\prime }}{\partial t}$.}:

\[
v=\sqrt{\left( {x^{\prime  \bullet}}\right) ^{2}+\left( %
{y^{\prime \bullet }}\right) ^{2}}
\]

then in thic case:

\[
v=\left\vert {z^{\prime \bullet }}%
\right\vert \mbox{.}
\]

Let function $z^{\prime }$ be a holomorphic function. Hence, in accordance
with the Cauchy-Riemann conditions the following equations are fulfilled:

\begin{eqnarray*}
\frac{\partial x^{\prime }}{\partial x} &=&\frac{\partial y^{\prime }}{%
\partial y}\mbox{,} \\
\frac{\partial x^{\prime }}{\partial y} &=&-\frac{\partial y^{\prime }}{%
\partial x}\mbox{.}
\end{eqnarray*}

Therefore, in accordance with (\ref{u7}):

\[
dz^{\prime }=e^{-\mathrm{i}\left( 2\alpha \right) }dz
\]

where $2\alpha $ is an holomorphic function, too. For example, let

\[
2\alpha :=\frac{1}{t}\left( \left( x+y\right) +\mathrm{i}\left(
y-x\right) \right) ^{2}\mbox{.}
\]

In this case:

\[
z^{\prime }=\int \exp \left( \frac{\left( \left( x+y\right) +i\left(
y-x\right) \right) ^{2}}{t}\right) dx+i\int \exp \left( \frac{\left( \left(
x+y\right) +i\left( y-x\right) \right) ^{2}}{t}\right) dy\mbox{.}
\]

Let $k:=y/x$.

Hence,

\[
z^{\prime }=\int \exp \left( \frac{\left( \left( x+kx\right) +i\left(
kx-x\right) \right) ^{2}}{t}\right) dx+i\int \exp \left( \frac{\left( \left( 
\frac{y}{k}+y\right) +i\left( y-\frac{y}{k}\right) \right) ^{2}}{t}\right) dy%
\mbox{.}
\]

Calculate:

\begin{eqnarray*}
&&\int \exp \left( \frac{\left( \left( x+kx\right) +i\left( kx-x\right)
\right) ^{2}}{t}\right) dx=\allowbreak \frac{1}{2}\sqrt{\pi }\frac{\mathrm{%
erf}\left( x\sqrt{-\frac{1}{t}\left( 2ik^{2}+4k-2i\right) }\right) }{\sqrt{-%
\frac{1}{t}\left( 2ik^{2}+4k-2i\right) }}\mbox{,} \\
&&i\int \exp \left( \frac{\left( \left( \frac{y}{k}+y\right) +i\left( y-%
\frac{y}{k}\right) \right) ^{2}}{t}\right) dy=\allowbreak \frac{1}{2}i\sqrt{%
\pi }\frac{\mathrm{erf}\left( y\sqrt{-\frac{1}{k^{2}t}\left(
2ik^{2}+4k-2i\right) }\right) }{\sqrt{-\frac{1}{k^{2}t}\left(
2ik^{2}+4k-2i\right) }}\mbox{.}
\end{eqnarray*}

Calculate:

\[
\frac{\partial z^{\prime }}{\partial t}=\frac{1}{-8\sqrt{t}i\left(
k-i\right) ^{3}\sqrt{-2i}}\allowbreak \left( 
\begin{array}{c}
-4y\left( k-i\right) ^{2}\sqrt{-\frac{1}{t}2i\left( k-i\right) ^{2}}\exp
\left( \frac{1}{k^{2}t}y^{2}2i\left( k-i\right) ^{2}\right)  \\ 
+4ikx\left( k-i\right) ^{2}\sqrt{-\frac{1}{k^{2}t}2i\left( k-i\right) ^{2}}%
\exp \left( \frac{1}{t}x^{2}2i\left( k-i\right) ^{2}\right)  \\ 
+i\sqrt{\pi }k^{2}t2i\left( k-i\right) ^{2}\sqrt{\frac{1}{t}}\sqrt{\frac{1}{%
k^{2}}}\mathrm{erf}\left( y\sqrt{-\frac{1}{k^{2}t}2i\left( k-i\right) ^{2}}%
\right)  \\ 
+\sqrt{\pi }kt2i\left( k-i\right) ^{2}\sqrt{\frac{1}{t^{2}}}\sqrt{\frac{1}{%
k^{2}}}\mathrm{erf}\left( x\sqrt{-\frac{1}{t}2i\left( k-i\right) ^{2}}\right) 
\end{array}%
\right) \mbox{.}
\]

For large $t$:

\[
\frac{\partial z^{\prime }}{\partial t}\approx\frac{1}{-8\sqrt{t}%
i\left( k-i\right) ^{3}\sqrt{-2i}}i\sqrt{\pi }k^{2}t2i\left( k-i\right) ^{2}%
\sqrt{\frac{1}{t}}\sqrt{\frac{1}{k^{2}}}\mathrm{erf}\left( y\sqrt{-\frac{1}{%
k^{2}t}2i\left( k-i\right) ^{2}}\right) 
\]

Hence,

\[
v\approx\left\vert \frac{1}{8}\left( 1-i\right) k\sqrt{\pi }\frac{1}{%
k-i}\mathrm{erf}\left( x\sqrt{-\frac{1}{t}2i\left( k-i\right) ^{2}}\right)
\right\vert \mbox{.}
\]

Because
\[
k=\tan \theta \mbox{, }x=r\cos \theta
\]
then
\[
v\approx \left\vert \frac{1}{8}\left( 1-i\right) \left( \tan \theta
\right) \sqrt{\pi }\frac{1}{\tan \theta -i}\mathrm{erf}\left( r\left( \cos
\theta \right) \sqrt{-\frac{1}{t}2i\left( \left( \tan \theta \right)
-i\right) ^{2}}\right) \right\vert 
\]%
$\mbox{.}$

Figure 6 shows the dependence of velocity $v$ on the radius $r$ at large $%
t\sim 10^4$ and at $\theta =13\pi/14$. Compare with Figure 4.

Figure 7 shows the dependence of velocity $v$ on the radius $r$ at large $%
t\sim 10^4$ and at $\theta =0.98\pi $. Compare with Figure 5.

\section{Conclusion}

Hence, Dark Matter and Dark Energy can be mirages in the space-time, which
is curved by oscillations of chromatic%
\index{chromatic} states.

\end{document}